\documentclass[twocolumn,aps,floats,letterpaper,floatfix,groupedaddress]{revtex4-1}
\usepackage{graphicx}
\usepackage{dcolumn}
\usepackage{amsmath}
\usepackage{amsfonts}
\usepackage{amssymb}
\usepackage{epsfig,float,afterpage,wrapfig,psfrag}
\usepackage{natbib}

\newcommand{\beq}{\begin{equation}}
\newcommand{\eeq}{\end{equation}}
\newcommand{\bes}{\begin{subequations}}
\newcommand{\ees}{\end{subequations}}
\newcommand{\bea}{\begin{eqnarray}}
\newcommand{\eea}{\end{eqnarray}}
\newcommand{\ba}{\begin{array}}
\newcommand{\ea}{\end{array}}
\newcommand{\beqn}{\begin{eqnarray*}}
\newcommand{\eeqn}{\end{eqnarray*}}

\newcommand{\f}[2]{\frac{#1}{#2}}

\newcommand{\dg}{\dagger}

\def\nn{\nonumber}

\newlength{\sizeonefig}
\newlength{\sizetwofig}
\begin{document}

\title{Topologically trivial zero bias conductance peak in semiconductor Majorana wires from boundary effects}

\author{Dibyendu Roy$^1$, Nilanjan Bondyopadhaya$^2$ and Sumanta Tewari$^3$}
\affiliation{$^1$ Theoretical Division and Center for Nonlinear Studies, Los Alamos National Laboratory, Los Alamos, NM 87545, USA}
\affiliation{$^2$ Integrated Science Education and Research Centre,  Visva-Bharati University, Santiniketan, WB 731235,  India}
\affiliation{$^3$ Department of Physics and Astronomy, Clemson University, Clemson, SC 29634, USA}

\begin{abstract}
We show that a topologically trivial zero bias conductance peak is produced in semiconductor-superconductor hybrid structures due to a suppressed superconducting pair potential and/or an excess Zeeman field at the ends of the heterostructure, both of which can occur in experiments. The zero bias peak (ZBP) (a) appears above a threshold parallel bulk Zeeman field, (b) is stable for a range of bulk field before splitting, (c) disappears with rotation of the bulk Zeeman field, and, (d) is robust to weak disorder fluctuations.
The topologically trivial ZBPs are also expected to produce splitting oscillations with the applied field similar to those from Majorana fermions. Because of such strong similarity with the phenomenology expected from Majorana fermions we find that the only unambiguous way to distinguish these trivial ZBPs (of height $4e^2/h$) from those arising from Majorana fermions (of height $2e^2/h$) is by comparing the (zero temperature) peak height and/or through an interference experiment.

\end{abstract}
\vspace{0.0cm}

\maketitle
The search for zero-energy Majorana fermions \cite{Majorana,Read00,Kitaev00,Wilczek} in solid state systems has received tremendous interest \cite{Nayak,Beenakker,Alicea12,Stanescu-Tewari} in recent theoretical and experimental studies.
A practical route to engineer such a system in the laboratory has been suggested in Refs. \cite{Sau10,Long-PRB,Lutchyn10,Oreg10}, following earlier similar proposals in topological insulators \cite{Fu08} and cold fermion systems \cite{Zhang08,Sato09}. The proposed system consists of a spin-orbit coupled semiconductor thin film or nanowire proximity coupled to a $s$-wave superconductor and in the presence of a suitably directed Zeeman spin splitting. The applied Zeeman field drives the engineered hybrid nanowire (semiconductor Majorana wire) through a topological quantum phase transition (TQPT) to a topologically nontrivial superconducting phase with localized zero energy Majorana bound states (MBSs) when the Zeeman splitting $\Gamma$ exceeds a critical value $\Gamma_c$.
 The Majorana wire is in a topologically trivial superconducting phase with no defect-localized MBS when the Zeeman splitting satisfies $\Gamma < \Gamma_c$. The zero energy MBS localized at the ends of the nanowire (for $\Gamma > \Gamma_c$) has been proposed to be observable by a zero bias conductance peak in charge current through the ends of the semiconductor wire \cite{Sengupta2001, Bolech07, Tewari08, Long-PRB, Law2009, Flensberg2010,Wimmer2011,Stanescu2011,Qu2011,Fidkowski2012, Mourik12,Deng12,Das12}. Here we show that such a charge current zero bias peak (ZBP) can exist even in the topologically trivial phase ($\Gamma < \Gamma_c$) due to excess Zeeman field and/or suppressed superconducting pair potential at the wire ends, both of which can occur in experiments. Our calculations establish that a (non-quantized) ZBP from tunneling into the ends of a spin-orbit coupled wire is unable to produce an unambiguous signature of MBS, \textit{even after taking into account} all the proposed experimental checks for Majorana fermions that have appeared in recent theoretical and experimental works \cite{Mourik12,Das12,Stanescu12,DasSarma-Sau-Stanescu}.

The critical Zeeman field $\Gamma_c$ $(= \sqrt{\Delta^2 + \mu^2})$ in the quasi-1D heterostructure depends on the value of the superconducting pair potential $\Delta$ proximity induced in the semiconductor and the chemical potential $\mu$ of the semiconductor measured from the top-most occupied confinement energy band (by ``band", we mean a pair of spin--split sub--bands, the band themselves being separated by the energy gaps due to lateral confinement).
The value of $\Gamma_c$ (for $\Delta \rightarrow 0$) corresponds to the value above which ($\Gamma > \mu$) an odd number of confinement induced sub-bands are occupied in the semiconductor.
For an even number of sub-band occupancy the system is in the topologically trivial phase with no MBS localized at the ends of the nanowire. 

The typical experimental nanowires, however, are more likely to have an even number of sub-band occupancy than odd. This is because the confinement energy gap $E_C$ is typically much larger than the Zeeman energy gap $\Gamma$, and, consequently,
the chemical potential is expected to lie in the confinement gap \textit{between} two pairs of spin--split sub-bands rather than in the Zeeman gap between two sub-bands in the same band. By a simple estimate (within the non-interacting electron model), assuming that the 1D density of states $\propto 1/\sqrt{E}$, the ratio of the number of samples with an even sub-band occupancy to that with an odd sub-band occupancy
is $\sim \sqrt{E_C/\Gamma} \sim 6$. Here we have used typical values for $E_C \sim 6-7 $ meV and the applied $\Gamma \sim \Delta \sim 150-200 \mu$eV from Ref.~[\onlinecite{Mourik12}]. It follows that, unless $\mu$ can be tuned by an externally applied gate potential (which is hard because of the strong coupling to a superconductor) the chemical potential in the experimental nanowires may cross both sub-bands of the top occupied band and the system may be topologically trivial. Here we show that, even in this case, a robust but topologically trivial ZBP can still appear in tunneling experiments due to excess local Zeeman field and/or suppressed superconducting pair potential at the wire ends. The excess Zeeman field at the wire ends can be due to the Meissner effect (the edge magnetic field larger than that in the bulk), or due to the presence of a magnetic impurity at the nanowire-lead interface, and
   the suppressed $\Delta$ at the ends of the wire can be a consequence of the (spatially varying) proximity effect near the boundary of the bulk superconductor. Thus, both effects, capable of producing a ZBP even in the topologically trivial phase, can occur in the experimental systems.
 %Remarkably, the topologically trivial ZBP satisfies the same constraints as those produced by Majorana fermions, with the one important difference that the peak we find is of height $4e^2/h$ while a Majorana peak should have a height half as much. Our results establish that it is difficult to identify a (non-quantized) ZBP in semiconductor Majorana wires with topological Majorana fermions unless it can be supported by a determination of the (zero temperature) peak height and/or by an interference experiment \cite{Sau-Swingle-Tewari}.
 
 About the origin of the topologically trivial peaks in our work we speculate on the following scenario: As discussed in Ref.~[\onlinecite{Kells12b}] the zero energy bound states from the two different sub-bands (in the top occupied band) at the same end of the nanowire could contribute to a total ZBP height of $4e^2/h$ \textit{provided their coupling and the resultant energy splitting could be suppressed}. Ref.~[\onlinecite{Kells12b}] proposed that a large uniform bulk Zeeman field, coupled with a smooth barrier potential at the lead-nanowire interface, can suppress the inter-sub-band s--wave pairing \cite{Alicea}, resulting in a near-zero-energy peak even in the topologically trivial phase. We speculate that the coupling between the two local zero energy states can be suppressed even when a perturbation 
 %(in the form of a variation of the Zeeman field and/or the superconducting pair potential) 
 is applied only near the heterostructure ends because the zero energy states from the (uncoupled) sub-bands are localized there within a length scale $\sim \xi$, the coherence length. The local excess Zeeman field and/or suppressed pair potential at the nanowire ends can however be a natural consequence of the experimental geometry and thus present an alternative explanation of the (unquantized) ZBPs seen in the recent experiments on semiconductor-superconductor heterostructures \cite{Mourik12,Deng12,Das12}

 The trivial ZBP appears in our calculations entirely in the topologically trivial phase, that is, with only an even number of confinement induced sub-bands occupied. According to the scenario presented above, even though the ZBP (of height $4e^2/h$) can be viewed as the result of resonant local Andreev reflection from two (or an even number of) weakly coupled MBSs, it is important to mention that no TQPT is necessary for the near-zero-energy states to appear. In other words, an even number of Majorana fermions, even if they are only weakly coupled, can be thought of as a conventional Dirac fermion, and do not have non-Abelian statistics. It is in this sense that we call the ZBPs observed with local perturbations in our work topologically trivial ZBPs.

We consider a one-dimensional multichannel Rashba spin-orbit coupled semiconducting nanowire (for example, InSb) oriented along the $x$-direction, and in proximity to a $s$-wave superconductor (for example, NbTiN). The full hybrid structure is modeled by a discrete $N \times W$ square-lattice tight-binding model \cite{Potter11, Sticlet12}
\bea
&&H_{NW}=H_{\text KE}+H_{\text SO}+H_{\text Z}+H_{\text SC}, \label{NWHam}\\
&&H_{KE}=\sum_{{\bf r},{\bf d},\alpha}-t(c^{\dg}_{{\bf r}+{\bf d},\alpha}c_{{\bf r},\alpha}+{\rm H.c.})+\tilde{\mu}(c^{\dg}_{{\bf r},\alpha}c_{{\bf r},\alpha}-\f{1}{2}),\nn\\
&&H_{\text SO}=\sum_{{\bf r},{\bf d},\alpha,\beta}-i\alpha_R c^{\dg}_{{\bf r}+{\bf d},\alpha} \hat{\bf z}.(\boldsymbol{\sigma}_{\alpha,\beta} \times {\bf d})c_{{\bf r},\beta}+{\rm H.c.},\nn\\
&&H_{\text Z}=\sum_{{\bf r},\alpha,\beta} c^{\dg}_{{\bf r},\alpha}({\bf B}.\boldsymbol{\sigma})_{\alpha,\beta}c_{{\bf r},\beta},H_{\text SC}=\sum_{{\bf r}}-\Delta c^{\dg}_{{\bf r},\uparrow}c^{\dg}_{{\bf r},\downarrow}+{\rm H.c.},\nn
\eea
where ${\bf r} \in \{l,m\}$ denotes lattice sites with the index $l$ along the length, $l=1,2..N$, and the index $m$ along the width of the nanowire, $m=1,2..W$, ($W$ counts the number of tight-binding chains) ${\bf d} \in \{\hat{e}_x,\hat{e}_y\}$ is a unit vector connecting nearest-neighboring sites, $(\alpha,\beta) \in \{\uparrow,\downarrow\}$ are spin indices, and $\boldsymbol{\sigma}$ is spin-$1/2$ Pauli matrix vector. Here $c^{\dg}_{{\bf r},\alpha}$ is a creation operator of an electron with spin $\alpha$ at lattice site ${\bf r}$ of the nanowire. The hopping strength $t$ of the electrons is related to the band mass $m^*=\hbar^2/(2ta^2)$ where $a$ denotes lattice spacing in the tight-binding model. %The Rashba spin-orbit coupling strength $\alpha_R$ can be very different in the semiconductor-superconductor hybrid structures from that in the bare semiconductor, thus it can be a source of uncertainty in the experiments. 
The applied magnetic field ${\bf B} \in \{B_x,B_y,B_z\}$ opens a Zeeman splitting $\Gamma = (g \mu_B/2) B$ in the sub-bands, and $\Delta$ is the proximity induced superconducting pair potential in the semiconductor. In the following we assume $g\mu_B/2 =1$ and identify $\Gamma$ with $B$ in the reduced units. In recent experiments \cite{Mourik12, Das12} ${\bf B}$ along the wire axis ($x$-axis) is increased to observe a ZBP, and then ${\bf B}$ is tilted from the wire axis to remove the ZBP, both observations consistent with the Majorana origin of the ZBP at the wire ends \cite{Long-PRB}. In Eq.~(\ref{NWHam}) $\tilde{\mu}=2(\mu-t)$ and an energy shift $-\tilde{\mu}/2$ has been added for the local Majorana transformation used in our transport calculations. We couple the semiconductor-superconductor hybrid structure in Eq.~(\ref{NWHam}) to two metallic leads at the two ends. The metallic leads are modeled by free electron tight binding chains. Each first $\{1,m\}$ and last sites $\{N,m\}$ with $m=1,2..W$ of all the transverse tight-binding chains of the semiconductor nanowire are coupled to a semi-infinite free-electron tight binding chain which forms the metallic leads. All the tight binding chains in the left lead are kept at chemical potential $\mu_L$ and temperature $T_L$, and those in the right lead are at chemical potential $\mu_R$ and temperature $T_R$. The Hamiltonian $H^p_{\text M}$ ($p = L, R$) below describe the metallic lead Hamiltonians and $H^p_{\text C}$ the corresponding tunnel couplings between the leads and the nanowire,
\bea
&& H_{\text M}^p = -\gamma_p \, \sum_{\alpha,m=1}^W \sum_{k =1}^{\infty}( c_{m, k, \alpha }^{p \dagger} \, c_{ m, k+1, \alpha}^p + c_{m, k+1, \alpha}^{p \dagger} \, c_{ m, k, \alpha}^p),  \nn \\
&& H^p_{\text C} = -\gamma_p' \, \sum_{\alpha, m=1}^W ( c_{m,1,\alpha}^{p \dagger}\, c_{m, l_p,\alpha} + c_{m, l_p,\alpha}^\dagger \, c_{m,1,\alpha}^p ).
\eea
where $l_{\text L} =1$ and $l_{\text R}=N$. Here $c^{p\dg}_{m,k,\alpha}$ is an electron creation operator on the $p$th lead. The strength of tunnel coupling between the normal leads and the nanowire is $\gamma_p'$. It controls the width and the height of the ZBP. Here we apply the quantum Langevin equations and Green's function (LEGF) approach \cite{DharSen06, DharRoy06, Roy07, Roy12, Roy13} to calculate the current-voltage $(I$-$V)$ characteristics  and the corresponding differential conductance $dI/dV$ across the multichannel spin-orbit coupled semiconductor nanowires. To mimic the recent experimental conditions \cite{Mourik12}, we set $\mu_R=0$ and drive $\mu_L$ from $-eV$ to $eV$. We fix the temperature of the two leads to be the same, $T_L=T_R=0$.

%{\it Methods:}
%First we transform the electron operators on the nanowire lattice in Eq.~(\ref{NWHam}) using a local Majorana transformation,
%\bea
%c_{l,m,\alpha}= \frac{1}{2} (C_{lA,m,\alpha} +\imath \, C_{lB,m,\alpha})
%\eea
%with $C^{\dagger}_{lA(B),m,\alpha}=C_{lA(B),m,\alpha}$ where $C_{lA(B),m,\alpha }$ is a local Majorana fermion operator. Next we apply quantum Langevin equations and Green's function (LEGF) method \cite{DharSen06, DharRoy06, Roy07} to find steady state solution of the Majorana operators on the nanowire after integrating out the lead operators. The LEGF method has been recently used to study the current-voltage $(I-V)$ characteristics of the 1D Kitaev chain (spinless p--wave superconductors) \cite{Roy12}, and it has been later extended to the single channel spin-orbit coupled semiconductor-superconductor hybrid structures \cite{Roy13}. Here we generalize the method for multichannel spin-orbit coupled semiconductor nanowires to derive the current-voltage characteristics. To mimic the recent experimental conditions \cite{Mourik12}, we set $\mu_R=0$ and drive $\mu_L$ from $-V$ to $V$. We fix the temperature of the two leads to be the same, $T_L=T_R=0$.

\begin{figure}
\includegraphics[width=8.0cm]{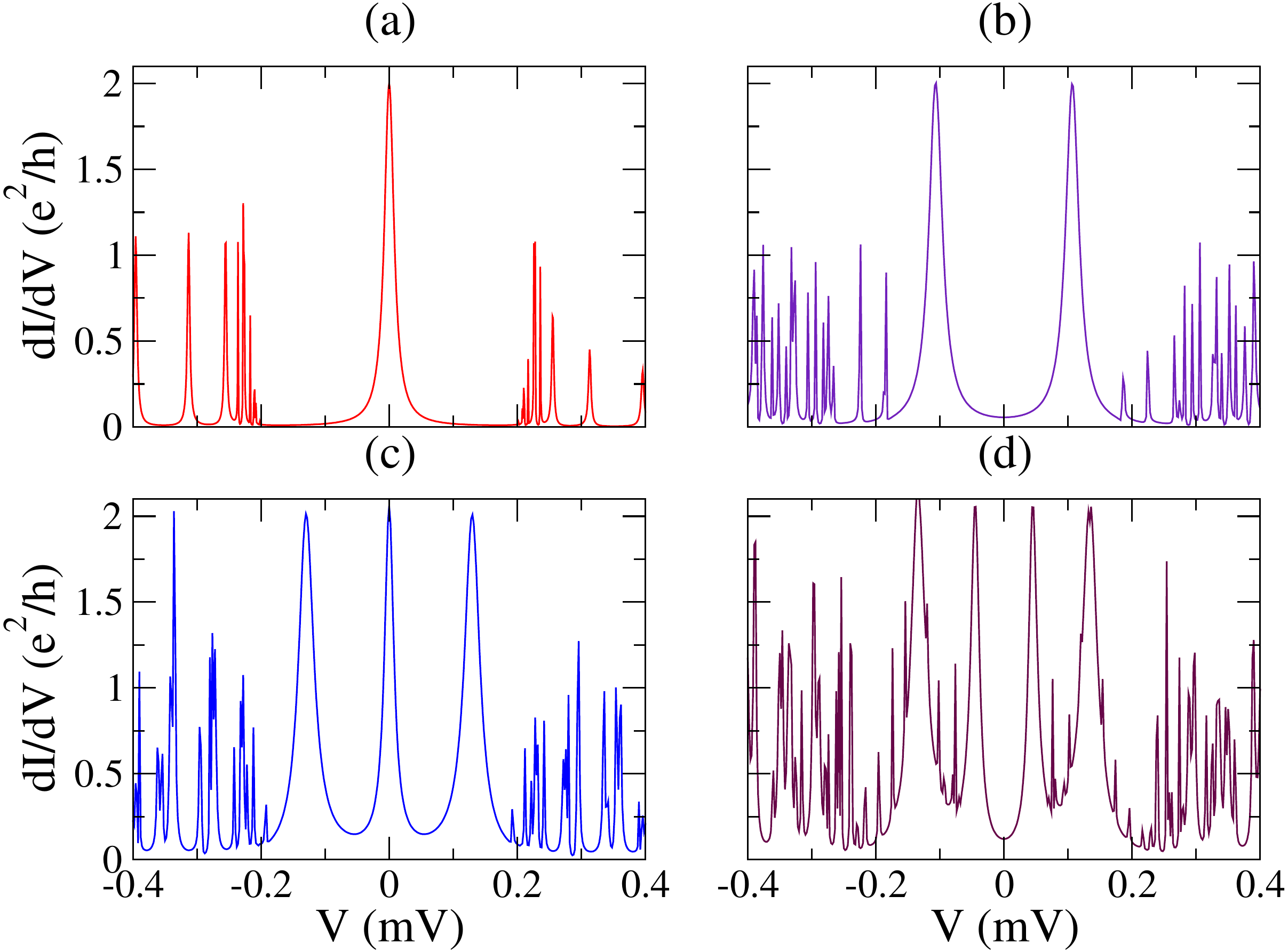}
\caption{Zero-temperature differential conductance $dI/dV$  \textit{vs.} the applied voltage $V$ for different numbers of transverse chains $W$ for $\mu=0$. Everywhere $B_x=0.8$, $B_y=B_z=0.0$ and $|\Delta|=0.6$, so $B_x > B_{x,c} = |\Delta|$. Number of chains $W$ in the panels is as follows: (a) $W=1$, (b) $W=2$, (c) $W=3$, and (d) $W=4$. Since $\mu=0$ in Eq.~(\ref{NWHam}) (with periodic boundary conditions) corresponds to an equal number of sub-bands below and above $E=\mu$, only $W=1,3$ has a ZBP for $B_x > B_{x,c}=|\Delta|$. For $W=2,4$, $B_x > |\Delta| $ produces no ZBP because only an even number of sub-bands are occupied.}
\label{ZBCP1}
\end{figure}

\begin{figure}
\includegraphics[width=8.0cm]{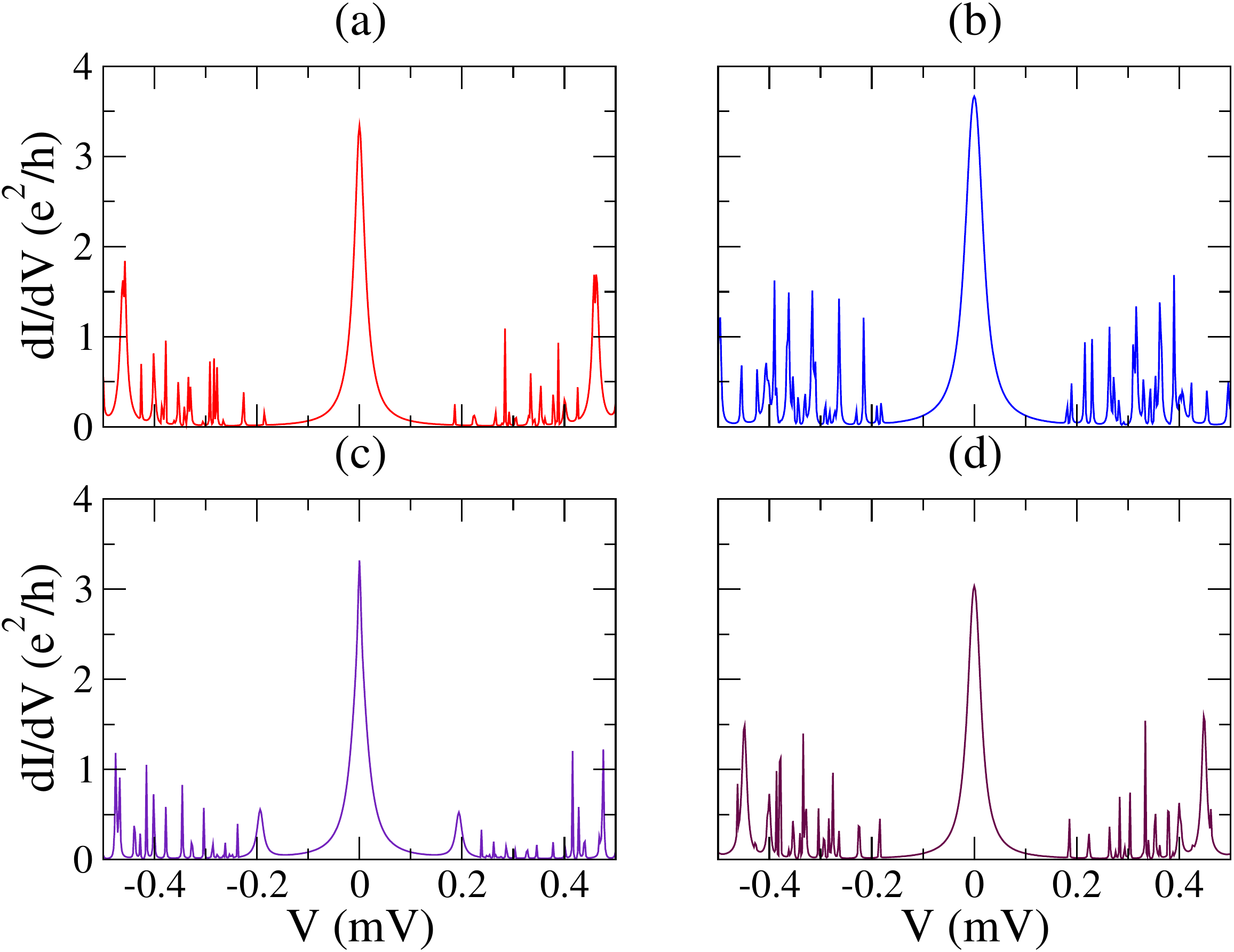}
\caption{Emergence of ZBP in the topologically trivial superconducting phase (even band occupancy) due to excess Zeeman field and/or reduced pair potential at the wire ends. In all the panels, $W=2, \Delta=0.6$, and $\mu=0$. In panels (a,b,d) $\alpha_R=0.2, B_x=0.8$. Note that for these values of $B_x$ and $\Delta$ there is no ZBP for $W=2$ (even band occupancy) in the absence of a local perturbation (Fig.~\ref{ZBCP1}b). The local end perturbations producing the topologically trivial ZBPs are as follows: (a) $\delta B_x\{1,m\}=0.45$, (b) $\delta\mu_{\uparrow}\{1,m\}=1,~\delta\mu_{\downarrow}\{1,m\}=0.2$, and (d) $\delta\Delta\{1,m\}=-0.35$, where $m=1,2$. In panel (c) we show, for illustrative purposes, the emergence of the trivial ZBP even for a larger value of the spin-orbit coupling, $\alpha_R=0.5$ and $B_x=1$. For local end perturbations we have used, $\delta B_x\{1,m\}=0.7, \delta B_x\{2,m\}=0.5$, that is, the end perturbation is applied on two sites and gradually peters of towards the bulk.}
\label{ZBCP2}
\end{figure}

An applied magnetic field along $x$ splits the two sub-bands of a single tight binding chain, and the MBSs appear when only one sub-band is occupied, i.e., the chemical potential lies in the gap between the two split sub-bands. The critical magnetic field along $x$ direction is given by $B_c=\sqrt{\Delta^2+\mu^2}$, the MBSs appear for $B_x>B_c$. The above formula for $B_c$ holds even for multichannel nanowires, however the chemical potential $\mu$ is determined from the bottom of the topmost occupied band. In our Hamiltonian (Eq.~(\ref{NWHam})) $\mu=0$ lies in the middle of all the sub-bands, and thus an odd number of sub-bands are occupied (for $B_x>\mu$) when the number of chains $W$ is odd, and an even number of sub-bands are occupied (for $B_x>\mu$) when $W$ is even. Therefore, a TQPT from a trivial superconducting phase to a topological superconductor occurs by increasing $B_x$ above $B_c$ for an odd $W$ but not for an even $W$ (for $\mu=0$). We use Fig.~\ref{ZBCP1} to illustrate this: there is a ZBP from the MBS for $W=1,3$ and no ZBP in the topologically trivial phase for $W=2,4$. We use $N=40$, $\gamma_{p}=t=1$, $\gamma^{\prime}_{p}=0.2$ ($p=L,R$), $\mu=0$, $\alpha_R=0.2$, $B_x=0.8$, $B_y=B_z=0$ and $|\Delta|=0.6$ in all the figures in this paper, if not explicitly stated otherwise. In physical units these parameters for a 2 $\mu m$ long wire correspond to a Rashba spin-orbit coupling (in the continuum model) $\alpha=\alpha_R a= 0.1~{\rm eV}\mathring{A}$, pair potential $\Delta=0.3$ meV and Zeeman splitting $\Gamma=0.4$ meV.
We have confirmed that the results in this paper are robust to variations in the values of these parameters in particular to the Rashba spin-orbit coupling parameter $\alpha$ (Fig.~\ref{ZBCP2}c). In all the figures we quote the parameters in units of $t=1$.

  Now we turn on a local perturbation to the magnetic field at the first sites $\{1,m\}$ ($m=1,2..W$) of all the transverse chains of the semiconductor nanowire. We find that a ZBP appears in the differential conductance of the nanowire \textit{even in the topologically trivial phase} (even number of occupied sub-bands for $\Delta \rightarrow 0$) when the magnitude of $B_x$ locally at $\{1,m\}$ is larger than that in the bulk (Fig.~2a). The ZBP also appears in the topologically trivial phase when there is a local $B_z$ field at the first sites (in addition to bulk $B_x$) of the same order of $B_x$ (Fig.~\ref{ZBCP2}b). Note that a local $B_z$ field is equivalent to a local shift of the chemical potential of the electrons with one spin with respect to the other. A topologically trivial ZBP appears also for a local suppression in the superconducting pair potential $\Delta$ at the first sites of the transverse chains (Fig.~\ref{ZBCP2}d). The heights of the trivial ZBPs are of the order $4e^2/h$ (Fig.~\ref{ZBCP2}), which is double the height produced by the MBSs (Fig.~\ref{ZBCP1}).

\begin{figure}
\includegraphics[width=8.0cm]{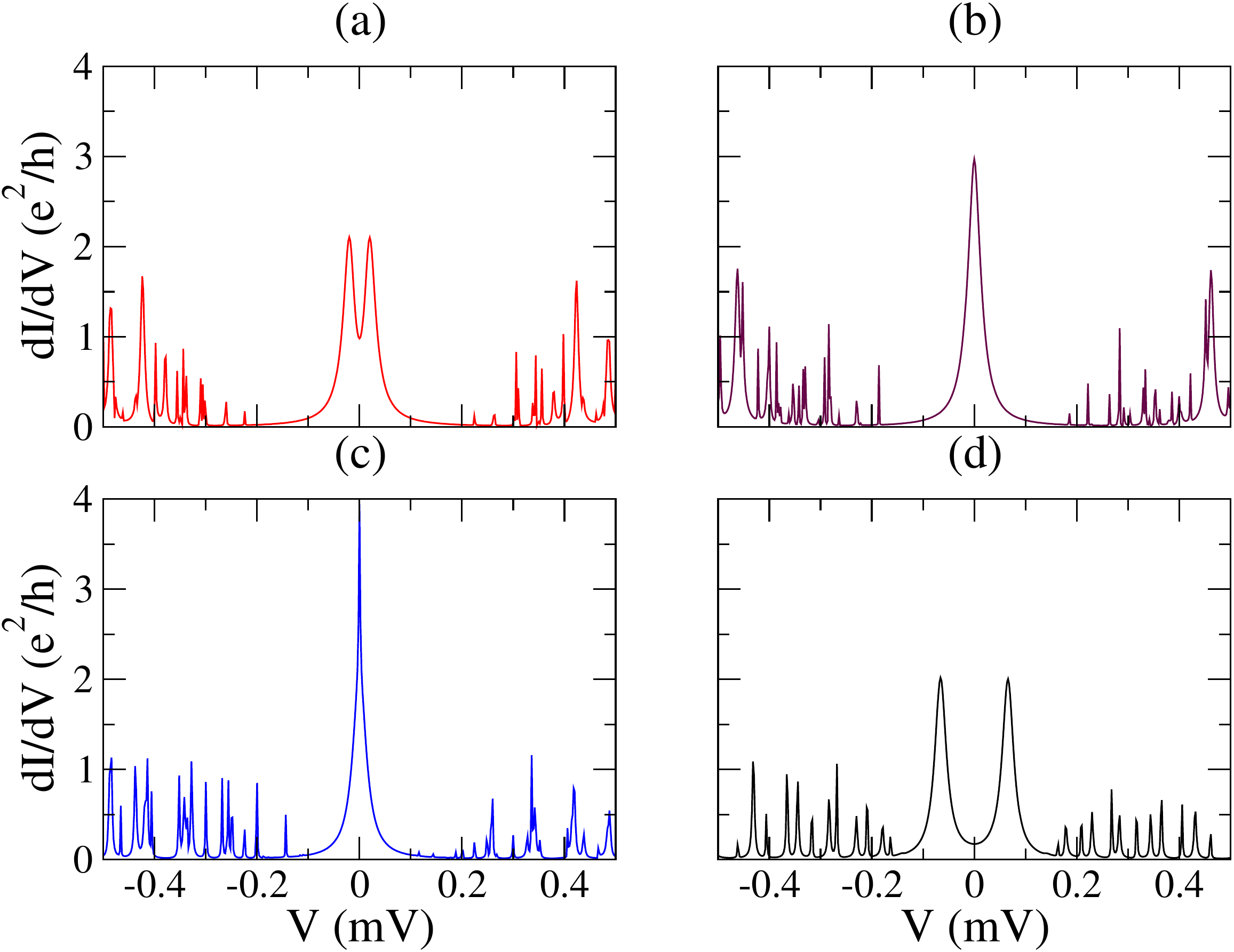}
\caption{Dependence of the topologically trivial ZBP on the uniform Zeeman field $B_x$ for a fixed perturbation at two end sites on all chains ($\delta B_x\{1,m\}=0.32$ and $\delta B_x\{2,m\}=0.2$). All panels have $W=2$ (two transverse chains) and $\Delta=0.6$. The ZBP is split in panel (a) ($B_x=0.7$), is fully developed in panel (b) ($B_x=0.8$), and stays in panel (c) ($B_x=1.0$), before splitting again with larger $B_x$ (not shown). Panel (d) ($B_x=0.4$, $B_y=0.4$) shows the disappearance of the ZBP with rotation of the uniform field in the plane formed by the wire and the effective spin-orbit coupling ($(x-y)$ plane).}
\label{ZBCP3}
\end{figure}

\begin{figure}
\includegraphics[width=8.0cm]{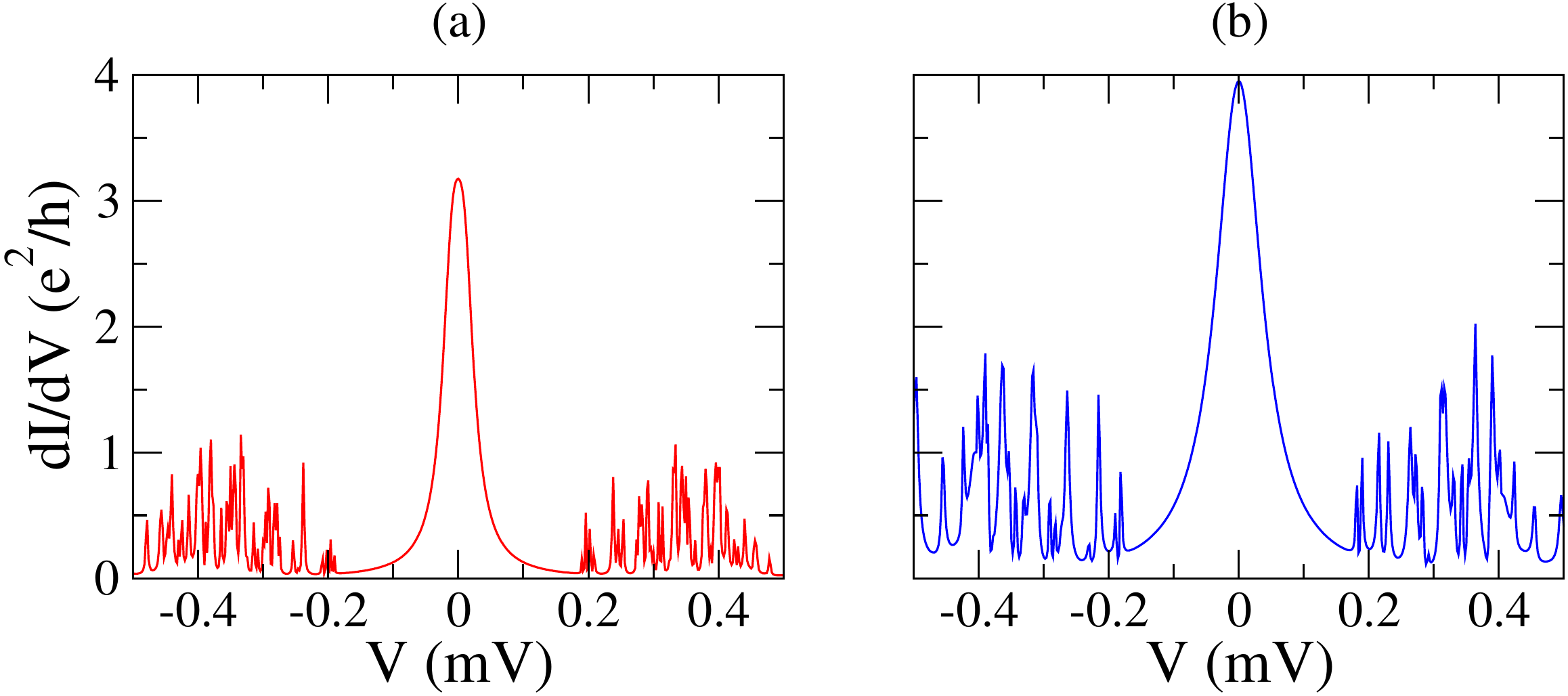}
\caption{Robustness of the topologically trivial ZBP to moderate disorder fluctuations and the role of barrier potential (tunnel coupling with the metallic lead). In all panels $W=2$, $B_x=0.8$, and $\mu_{\uparrow}\{1,m\}=1,~\mu_{\downarrow}\{1,m\}=0.2$. (a) Randomness in $\mu$ is chosen from a uniform distribution between 0 and 0.1, $\gamma^{\prime}_{p}=0.2$ and (b) clean wire ($\mu=0$), but a stronger coupling with the lead, $\gamma^{\prime}_{p}=0.3$ ($p=L,R$). These figures should be compared with that in Fig.~\ref{ZBCP2}b.}
\label{ZBCP4}
\end{figure}

 Applying the local perturbation at multiple boundary sites on all the chains (Fig.~\ref{ZBCP2}c) we find that the topologically trivial ZBP appears in a larger range of the local perturbations than when the perturbation is applied at a single boundary site (as in Fig.~\ref{ZBCP2}a,\ref{ZBCP2}b,\ref{ZBCP2}d). This is also more natural experimentally since the excess Zeeman field and/or the suppressed pair potential at the wire ends are expected to gradually shift to the bulk values in the interior of the nanowire.
 In Fig.~\ref{ZBCP3}a,\ref{ZBCP3}b we show that the trivial ZBPs due to a fixed set of local perturbations in the Zeeman field (applied at the two end sites) appear only above a threshold uniform Zeeman field in the bulk. The threshold bulk Zeeman field itself depends on the value of the local perturbations in the magnetic field or the pairing potential. For example, we find that the threshold field required for the emergence of the ZBP decreases with a larger value of the local perturbation. In Fig.~\ref{ZBCP3}c we show that the ZBP is robust to a further increase in the bulk field, before splitting into two peaks above a higher threshold (not shown). This sort of dependence of the topologically trivial ZBP on the uniform bulk field is consistent with the recent experiments \cite{Mourik12,Deng12,Das12}. The splitting of the trivial ZBP with increasing magnetic field can arise from two mechanisms. First, due to the overlap between the nearly zero energy states from the two ends of the nanowire, and second due to the local overlap between the zero energy states at the same end of the nanowire.
  In Fig.~\ref{ZBCP3}d we show that the trivial ZBP disappears as the bulk Zeeman field is rotated by $\pi/4$ from the axis of the wire in the plane formed by the wire axis and the direction of the effective spin-orbit field (the $(x-y$) plane). This behavior is also fully consistent with the recent experiments \cite{Mourik12,Das12}, which is taken as an argument in favor of the Majorana origin of the ZBP \cite{Long-PRB}.

We have checked the robustness of the topologically trivial ZBP to a moderate and random impurity distribution chosen independently for each site, identically distributed according to a uniform distribution. We find that the trivial ZBP survives even in the presence of moderate disorder fluctuations as shown in Fig.~\ref{ZBCP4}a. With respect to the clean wire, the height (width) of the trivial ZBP decreases (increases) with disorder compared to that without disorder in Fig.~\ref{ZBCP2}b.
However the trivial ZBP splits into two peaks for further increase of the strength or randomness of the impurity potential.
For the dependence of the ZBP on the barrier potential at the lead-wire interface \cite{Kells12b,Stanescu12}, which, in our calculations, controls the coupling $\gamma^{\prime}_p$ (see Eq.~(2)) of the nanowire to the metallic leads, we find that the height and width of the ZBP are somewhat susceptible to the barrier potential.
For weaker coupling to the lead, the height and width of the peak are reduced, while they increase with increasing the coupling to the lead as shown in Fig.~\ref{ZBCP4}b. For a weaker coupling to the leads (higher barrier potential) we find that the range of local perturbations in which the trivial ZBP is formed is reduced compared to the case with a stronger lead coupling (lower or smoother barrier potential). For the so-called gap-closing signature \cite{Mourik12,Stanescu12,Stanescu-Tewari-Sau} before the appearance of the ZBP, note that no gap needs to close before the appearance of the topologically trivial ZBP because the system remains in the same topologically trivial superconducting phase. This is one important difference between the Majorana origin of the ZBP and the topologically trivial ZBP as found here. Even if there is no bulk gap closing involved (no TQPT), for a linear dependence of the end perturbations on the bulk parameters, the spectral weight should continuously shift from higher subgap energies to zero energy as the uniform bulk field crosses the threshold (Fig.~\ref{ZBCP3}a, \ref{ZBCP3}b)). This may appear as a gap closing signature similar to the calculations in Ref.~[\onlinecite{Stanescu12},\onlinecite{Liu12}]. However, for a non-linear dependence of the local perturbations on the bulk parameters, the downward shift of the spectral weight may not be gradual with the uniform field, in which case any gap closing signature may be suppressed.

In summary, we investigate the effects of an excess Zeeman field and/or suppressed pair potential at the ends of the semiconductor Majorana wire on the tunneling conductance through the wire ends. Both effects may be present in the experiments and we show that both can produce stable ZBPs above a threshold uniform bulk Zeeman field, \textit{even when the system is topologically trivial} (sub-band occupancy even). 
%The excess Zeeman field at the nanowire ends can be a result of the Meissner effect and/or local magnetic impurities while the suppressed pair potential at the ends can be due to non-uniform proximity effect near the ends of the bulk superconductor.
%We find that such local perturbations can induce near-zero-energy bound states at the wire ends and a resultant ZBP in the tunneling measurements even when the semiconducting wire is in the topologically trivial phase. The ZBP appears only after a threshold uniform bulk Zeeman field and is stable for a reasonable range of the bulk field before splitting. The peak is removed when the bulk field is rotated in space, consistent with the experiments \cite{Mourik12,Das12}.
%The topologically trivial ZBPs are also robust to weak disorder fluctuations, and, depending on the correlation between the end and the bulk Zeeman fields, the shift in the spectral weight from higher energy to the zero bias as a function of the bulk field (the so-called ``gap closing signature") may or may not be visible in the experiments. 
The trivial ZBPs, since they arise from conventional BdG states, are expected to show splitting oscillations with the bulk field \cite{DasSarma-Sau-Stanescu} similar to the Majorana fermions.
Given such strong similarities with the ZBPs observable from the MBSs, the only unambiguous way to distinguish the trivial ZBPs found here (of height $4e^2/h$) from those arising from Majorana fermions (of height $2e^2/h$) is by comparing the (zero temperature) peak height and/or by an interference experiment \cite{Sau-Swingle-Tewari}.

DR acknowledges support of the U.S. Department of Energy through LANL/LDRD Program and ST acknowledges DARPA-MTO (FA9550-10-1-0497) and NSF (PHY-1104527) for support.


\begin{thebibliography}{34}
\expandafter\ifx\csname natexlab\endcsname\relax\def\natexlab#1{#1}\fi
\expandafter\ifx\csname bibnamefont\endcsname\relax
\def\bibnamefont#1{#1}\fi
\expandafter\ifx\csname bibfnamefont\endcsname\relax
\def\bibfnamefont#1{#1}\fi
\expandafter\ifx\csname citenamefont\endcsname\relax
\def\citenamefont#1{#1}\fi
\expandafter\ifx\csname url\endcsname\relax
\def\url#1{\texttt{#1}}\fi
\expandafter\ifx\csname urlprefix\endcsname\relax\def\urlprefix{URL }\fi
\providecommand{\bibinfo}[2]{#2}
\providecommand{\eprint}[2][]{\url{#2}}

\bibitem{Majorana} E. Majorana, Nuovo Cimento \textbf{14}, 171 (1937).

\bibitem{Read00} N. Read and D. Green, Phys. Rev. B {\bf 61}, 10267 (2000).

\bibitem{Kitaev00} A. Yu Kitaev, Physics-Uspekhi {\bf 44}, 131 (2001).

\bibitem{Wilczek} F. Wilczek, Nature Physics \textbf{5}, 614 (2009).

\bibitem{Nayak} C. Nayak, S. H. Simon, A. Stern, M. Freedman, S. Das Sarma,  Rev. Mod. Phys. \textbf{80}, 1083 (2008).

\bibitem{Beenakker} C. W. J. Beenakker, arXiv:1112.1950 (2011).

\bibitem{Alicea12} J. Alicea, Rep. Prog. Phys. \textbf{75}, 076501 (2012).

\bibitem{Stanescu-Tewari} T. D. Stanescu, S. Tewari, arXiv:1302.5433 (2013).

\bibitem{Sau10} J. D. Sau, R. M. Lutchyn, S. Tewari, and S. Das Sarma, Phys. Rev. Lett. {\bf 104}, 040502 (2010).


\bibitem{Long-PRB} J. D. Sau, S. Tewari, R. Lutchyn, T. Stanescu and S. Das Sarma, Phys. Rev. B \textbf{82}, 214509 (2010).


\bibitem{Lutchyn10} R. M. Lutchyn, J. D. Sau, and S. Das Sarma, Phys. Rev. Lett. {\bf 105},  077001 (2010).

\bibitem{Oreg10} Y. Oreg, G. Refael, and F. von Oppen,  Phys. Rev. Lett. {\bf 105}, 177002 (2010).


\bibitem{Fu08} L. Fu and C. L. Kane, Phys. Rev. Lett. {\bf 100}, 096407 (2008).

\bibitem{Zhang08} C. Zhang, S. Tewari, R. M. Lutchyn and S. Das Sarma, Phys. Rev. Lett {\bf 101}, 160401 (2008).

\bibitem{Sato09} M. Sato, Y. Takahashi, S. Fujimoto,  Phys. Rev. Lett {\bf 103}, 020401 (2009).

\bibitem{Sengupta2001} K. Sengupta, I Zutic, H. -J. Kwon, V. M. Yakovenko, S Das Sarma, 	Phys. Rev. B {\bf 63}, 144531 (2001).

\bibitem {Bolech07} C. J. Bolech and E. Demler, Phys. Rev. Lett. \textbf{98},
237002 (2007).

\bibitem{Tewari08} S. Tewari, C. Zhang, S. Das Sarma, C. Nayak, and D. -H. Lee, Phys. Rev. Lett. {\bf 100}, 027001 (2008).


%\bibitem{Tewari2008} S. Tewari, C. Zhang, S. Das Sarma, C. Nayak, D. -H. Lee, 	 New Journal of Physics {\bf 100} 027001 (2008).

\bibitem{Law2009} K. T. Law, P. A. Lee, T. K. Ng, Phys. Rev. Lett. {\bf 103} 237001 (2009).

\bibitem{Flensberg2010} K. Flensberg, Phys. Rev. B {\bf 82} 180516 (R) (2010).

\bibitem{Wimmer2011} 	M. Wimmer, A. R. Akhmerov, J. P. Dahlhaus, C. W. J. Beenakker, New Journal of Physics {\bf 13} 053016 (2011).

\bibitem{Stanescu2011} 	T. D. Stanescu, R. M. Lutchyn, S. Das Sarma, Phys. Rev. B {\bf 84} 144522 (2011).

\bibitem{Qu2011} C. Qu, Y. Zhang, L. Mao, C. Zhang, arXiv:1109.4108 (2011).


\bibitem{Fidkowski2012} L. Fidkowski, J. Alicea, N. Lindner, R. M. Lutchyn, M. P. A. Fisher, 	Phys. Rev. B {\bf 85} 245121 (2012).


\bibitem{Mourik12} V. Mourik, K. Zuo, S. M. Frolov, S. R. Plissard, E. P.
A. M. Bakkers and L. P. Kouwenhoven, Science {\bf 336}, 1003 (2012).

\bibitem{Deng12} M. T. Deng, C. L. Yu, G. Y. Huang, M. Larsson, P.
Caroff, H. Q. Xu, Nano Lett. {\bf 12}, 6414 (2012).

\bibitem{Das12} A. Das, Y. Ronen, Y. Most, Y. Oreg, M. Heiblum, H.
Shtrikman, Nat. Phys. {\bf 8}, 887 (2012).

\bibitem{Stanescu12} T. D. Stanescu and S. Tewari, arXiv:1208.6298 (2012).

\bibitem{DasSarma-Sau-Stanescu} S. Das Sarma, J. D. Sau, T. D. Stanescu, Phys. Rev. B \textbf{86}, 220506 (R) (2012).

\bibitem{Kells12b} G. Kells, D. Meidan, P. W. Brouwer, Phys. Rev. B {\bf 86}, 100503 (R) (2012).

\bibitem{Alicea} J. Alicea, Phys. Rev. B \textbf{81}, 125318 (2010).

%\bibitem{Sau-Swingle-Tewari} J. D. Sau, B. Swingle, S. Tewari, arXiv:1210.5514 (2012).


%\bibitem{Kells12b} G. Kells, D. Meidan, P. W. Brouwer, Phys. Rev. B {\bf 86}, 100503 (R) (2012).

%\bibitem{Alicea} J. Alicea, Phys. Rev. B \textbf{81}, 125318 (2010).


\bibitem{Potter11} A. C. Potter and P. A. Lee, Phys. Rev. B {\bf 83}, 094525 (2011).

\bibitem{Sticlet12} D. Sticlet, C. Bena, and P. Simon, Phys. Rev. Lett. {\bf 108}, 096802 (2012).

\bibitem{DharSen06} A. Dhar and D. Sen,  Phys. Rev. B {\bf 73}, 085119 (2006).

\bibitem{DharRoy06} A. Dhar and D. Roy, J. Stat. Phys. {\bf 125}, 801 (2006).

\bibitem{Roy07} D. Roy and A. Dhar, Phys. Rev. B {\bf 75}, 195110 (2007).

\bibitem{Roy12} D. Roy, C. J. Bolech, and N. Shah, Phys. Rev. B {\bf 86}, 094503 (2012).

\bibitem{Roy13} D. Roy, C. J. Bolech, and N. Shah, arXiv:1303.7036 (2013).

%\bibitem{Kells12b} G. Kells, D. Meidan, P. W. Brouwer, Phys. Rev. B {\bf 86}, 100503 (R) (2012).

%\bibitem{Alicea} J. Alicea, Phys. Rev. B \textbf{81}, 125318 (2010).

\bibitem{Stanescu-Tewari-Sau} T. D. Stanescu, S. Tewari, J. D. Sau, S Das Sarma, Phys. Rev. Lett. \textbf{109}, 266402 (2012).

\bibitem{Liu12} J. Liu, A. C. Potter, K. T. Law, P. A. Lee, Phys. Rev. Lett. {\bf 109}, 267002 (2012).

\bibitem{Sau-Swingle-Tewari} J. D. Sau, B. Swingle, S. Tewari, arXiv:1210.5514 (2012).

\end{thebibliography}
\end{document}